\documentclass[MSNbibl,nameyear,dvips]{arxstspdf}
\usepackage{dcolumn}
\usepackage{graphicx}
\usepackage{flushend}
\usepackage{stfloats}
\volume{29}
\issue{1}
\pubyear{2014}
\firstpage{81}
\lastpage{90}
\doi{10.1214/12-STS412} %kopijuoti is PTS
\makeatletter

\newcolumntype{d}[1]{D{.}{.}{#1}}

\def\Fbh{F_{\mathrm{bh}}}
\def\acool{\alpha_{\mathrm{cool}}}
\def\arh{\alpha_{\mathrm{reheat}}}
\def\alphahot{\alpha_{\mathrm{hot}}}
\def\vhotdisk{V_{\mathrm{hot}, \mathrm{disk}}}
\def\vhotburst{V_{\mathrm{hot}, \mathrm{burst}}}
\def\alphareheat{{\arh}}
\def\alphacool{{\acool}}
\def\epsilonStar{\varepsilon_{\star}}
\def\alphastar{\alpha_{\star}}
\def\taumrg{f_{\mathrm{df}}}
\def\VCUT{v_{\mathrm{cut}}}
\def\ZCUT{z_{\mathrm{cut}}}
\def\fellip{f_{\mathrm{ellip}}}
\def\fburst{f_{\mathrm{burst}}}
\def\FSMBH{{\Fbh}}
\def\yield{p_{\mathrm{yield}}}
\def\stabledisk{f_{\mathrm{stab}}}
\def\epsilonSMBHEddington{\varepsilon_{\mathrm{Edd}}}
\def\tdisk{t_{\mathrm{disk}}}

\makeatother

\begin{document}
\begin{frontmatter}

\title{Galaxy Formation: Bayesian History Matching for the Observable
Universe}%\thanksref{T1}
% kai straipsnis turi susijusiu diskusiju ir rejoinder'iu
%rejoinder at \relateddoi{r}{10.1214/00-STSXXXX}.}
\runtitle{Galaxy Formation: Bayesian History Matching}

\begin{aug}
\author[a]{\fnms{Ian} \snm{Vernon}\corref{}\ead[label=e1]{i.r.vernon@durham.ac.uk}},
\author[a]{\fnms{Michael} \snm{Goldstein}}
\and
\author[b]{\fnms{Richard} \snm{Bower}}
\runauthor{I. Vernon, M. Goldstein and R. Bower}

\affiliation{Durham University}

\address[a]{Ian Vernon is Lecturer and Michael Goldstein is Professor,
Department of Mathematical Sciences, Durham University, Science
Laboratories, South Road,
Durham, DH1 3LE, United Kingdom \printead{e1}.}
\address[b]{Richard Bower is Professor, ICC/Department of Physics, Durham
University, Science Laboratories, South Road,
Durham, DH1 3LE, United Kingdom.}

\end{aug}

% ABSTRACT
\begin{abstract}
Cosmologists at the Institute of Computational Cosmology, Durham
University, have developed a state of the art model of galaxy
formation known as Galform, intended to contribute to our
understanding of the formation, growth and subsequent evolution
of galaxies in the presence of dark matter.  Galform requires the
specification of many input parameters and takes a significant
time to complete one simulation, making comparison between the
model's output and real observations of the Universe extremely
challenging.  This paper concerns the analysis of this problem
using Bayesian emulation within an iterative history matching
strategy, and represents the most detailed uncertainty analysis
of a galaxy formation simulation yet performed.
\end{abstract}

% KEYWORDS

\end{frontmatter}

%s1 #&#
\section{Introduction}

Understanding the evolution of the universe from the Big Bang to
the current day is the fundamental goal of cosmology. A major
part of this is the problem of structure formation: understanding
the formation, growth and subsequent evolution of galaxies in the
presence of dark matter. The
world leading Galform group, based at the Institute of
Computational Cosmology, Durham University, has developed a state
of the art model of galaxy formation know as Galform. However,
they face a critical problem. Galform requires the specification
of many input parameters and takes a significant time to complete
one simulation, making comparison between the model's output and
real observations of the universe extremely challenging.

Here we describe the analysis of this problem using Bayesian
history matching methodology, highlighting why the problem itself
can only be sensibly formulated within a subjective Bayesian
context and demonstrating the use of Bayesian emulators within an
iterative history matching strategy. This work represents the
most detailed uncertainty analysis of a galaxy formation
simulation yet performed and, to our knowledge, the most
detailed history match, with the most number of iterations
completed, for any model in the scientific literature. This
methodology is widely applicable across any scientific discipline
that uses computer simulations of complex physical processes.

We discuss galaxy formation in Section~\ref{secRB}, the Bayesian
history matching methodology in Section~\ref{secMG} and the
application and results in Section~\ref{secIV}. For a more detailed
account of this ongoing project see \citet{Vernon10CS}.

%f1 #&#
\begin{figure*}

\includegraphics{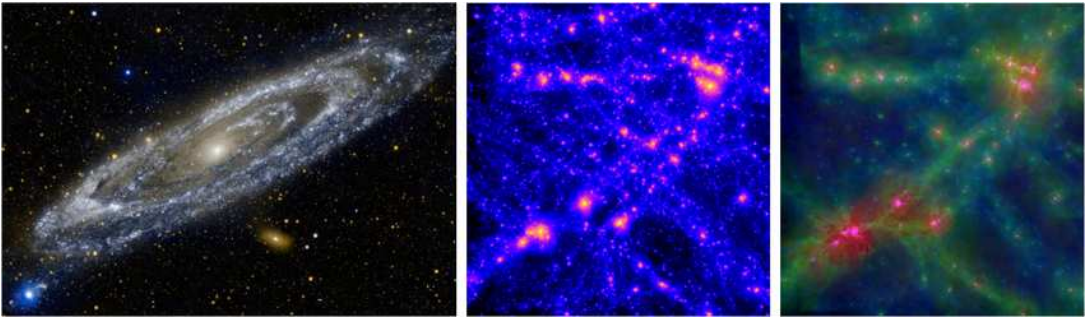}

\caption{Left: the Andromeda galaxy (NASA), the closest large galaxy to
our own, contains approximately 1 trillion stars. Typical output from a
galaxy formation simulation showing the configuration of dark matter
(middle) and baryonic stars and gas (right) (Eagle collaboration).}
\label{galpics}\vspace*{-3pt}
\end{figure*}

%s2 #&#
\section{Galaxy Formation}\label{secRB}

%s2.1 #&#
\subsection{A Universe Full of Galaxies}

The night sky is full of stars. Yet the stars that are visible to the human
eye are only an unimaginably tiny fraction of the stars in the universe
as a whole. Equipped with telescopes, we discover that at great distances
beyond our own galaxy lie millions of millions of other galaxies, each with
their own populations of stars.
Moreover, galaxies come in a great variety of shapes and forms. Our own
Milky Way
galaxy is one of the larger spiral type galaxies. Spiral galaxies are dominated
by a flat disk of stars, often with prominent spiral arms (Figure~\ref{galpics}).
With modern telescopes, it has become possible to study galaxies at
greater and
greater distances from earth. Because of the finite speed of light,
such distant galaxies
are seen when the universe was much younger. Astronomers can use this
time delay to observe
the buildup and formation of galaxies.

These observations have revealed some, at first sight, puzzling results.
Explaining the tension between
the prima facie theoretical expectation and the observational evidence
was one of the
key motivations for developing the theoretical model discussed below.
The problem for current theories of galaxy formation is
not so much to understand why galaxies form, but to understand why they are
relatively small and few.
The basic ingredients are clear (the force of gravity and radiative
cooling of baryonic matter), but
we are only now beginning to understand how the formation
of galaxies is regulated. The surprising result is that the black holes
(the densest
objects in the universe) appear to play a key role in this.

So how do galaxies form? Why is the universe filled with such objects?
In principle,
it is a straightforward consequence of the dominance of the
gravitational force.
Since all matter makes a positive contribution to the gravitational force,
the clumping of the universe's mass is a run away process. As the condensations
of matter become denser, they become more effective as
attractors.\vadjust{\goodbreak}
These matter concentrations are referred to as haloes.
The observational evidence shows that most of this mass, however,
is not normal, ``baryonic,'' matter (that you and I are made from) and that
the universe is dominated by ``Cold dark matter'' (CDM): massive
particles that interact
very weakly (possibly associated with super-symmetric extensions
of the standard model of particle physics).

The CDM particles explain the collapse and growth of the gravitating
dark matter haloes,
but to populate these haloes with
luminous galaxies, we must turn to the astrophysics of the baryonic matter.
As the baryons are pulled together by the collapse
of the dark matter halo, they heat up and start to resist further
compression. The baryonic gas (but not the
collision-less dark matter) radiates this energy and cools, leading to
a run-away
contraction that is only stopped by the conservation of angular momentum.
The baryons form a thin, cold spinning disk of gas. Further
condensation leads to
the formation of stars and black holes.
In this scenario, most haloes are able to convert almost all their
baryonic component into
stars, but this is in direct conflict with the observed 10\% baryonic conversion
The origin of this
discrepancy is a key cosmological puzzle and astronomers
appeal to ``feedback'' to resolve it: somehow the formation of stars
and black holes
must inject energy that prevents further gas cooling. One of the key
aims of
the Galform project is to identify the feedback schemes that are needed to
account for the observed universe.

%t1 #&#
\begin{table*}
\caption{Table of the 17 input parameters that make up the vector $x$
and associated ranges (which were converted to $-1$ to 1 for
the analysis). Input parameters are grouped by physical process}
\label{tabinputs}
\begin{tabular*}{\tablewidth}{@{\extracolsep{\fill}}ld{3.2}d{4.2}ccd{2.3}d{2.2}c@{}}
\hline
\textbf{Input} & & & \textbf{Process} & \textbf{Input}
&&&\textbf{Process} \\
\textbf{parameter} $\bolds{x}$
&\multicolumn{1}{c}{\textbf{Min}}&\multicolumn{1}{c}{\textbf{Max}} &
\textbf{modelled} & \textbf{parameter}
$\bolds{x}$ & \multicolumn{1}{c}{\textbf{Min}}
& \multicolumn{1}{c}{\textbf{Max}} & \textbf{modelled} \\
\hline
$\vhotdisk$ &100&550 & SNe feedback & $\alphacool$&0.2&1.2 & AGN
feedback\\
$\vhotburst$&100&550 & $\cdot$ & $\epsilonSMBHEddington$&0.004&0.05
& $\cdot$ \\
$\alphahot$&2&3.7 & $\cdot$ & $\taumrg$&0.8&2.7 & Galaxy mergers\\
$\alphareheat$&0.2&1.2 & $\cdot$ & $\fellip$&0.1&0.35 & $\cdot$ \\
$\epsilonStar$&10&1000 & Star formation & $\fburst$&0.01&0.15& $\cdot
$ \\
$\alphastar$&-3.2&-0.3& $\cdot$ & $\FSMBH$&0.001&0.01 & $\cdot$ \\
$\yield$&0.02&0.05 & $\cdot$ & $\VCUT$&20&50 & Reionisation\\
$\tdisk$& 0 & 1 & $\cdot$ & $\ZCUT$&6&9 & $\cdot$ \\
$\stabledisk$&0.65&0.95 & Disk stability & & & & \\
\hline
\end{tabular*}
\end{table*}

%s2.2 #&#
\subsection{Modeling Galaxy Formation with Galform}

Feedback greatly complicates an otherwise almost straightforward
problem. In order to
solve the problem from ab-initio principles, we would need to model the
formation
of individual stars and black holes.\vadjust{\goodbreak} Fortunately, we can make progress by
parameterising our lack of knowledge as uncertain coefficients in
formulae that summarise macroscopic
effects, and then by adjusting
these coefficients to provide the best description of the observed universe.
For example, although we cannot derive the rate of star formation from
the first principles, we
can include a parameter that describes the rate at which cold gas is
converted to stars and then
attempt to determine its plausible range of values through comparison
with observations.

The Galform code used in this project represents the state-of-the-art
in this approach.
It has been used to establish a very plausible model for the formation
of galaxies
(\cite{B06}) that describes many of the observed properties of the
galaxy population,
as diverse as the abundance of galaxies of different masses and the
history of the growth of
their black holes. It also makes well-tested predictions for properties
of the gas that
is left over from galaxies (\cite{Bower12Galform}). The model combines
many physical ingredients,
including modules to track: the gravitational collapse and buildup of
dark matter haloes; the cooling and accretion of gas;
the formation of stars, stellar evolution and ``feedback'' from
supernova explosions;
galaxy mergers and instabilities in stellar disks; the formation of
black holes and the associated feedback.
The modules link together to form a network of nonlinear equations that
are integrated in
time to trace the evolving properties of the galaxy population (see
Figure~\ref{galpics}).

%s2.3 #&#
\subsection{Galform Input and Output Parameters}

Each module has associated parameters. For example, these might specify
the rate at which cold gas is converted into stars, $\epsilonStar$, or
the energy generated in supernova feedback and its dependence on galaxy
mass $\vhotdisk$ and $\vhotburst$. Galform requires a total of 17 such
input parameters, shown in Table~\ref{tabinputs} along with appropriate
ranges elicited from the cosmologists and with the physical module each
parameter relates to. Exploring this 17-dimensional space is vital but
extremely challenging, as Galform takes approximately \textit{20 hours}
to complete a single evaluation. It also requires a detailed forcing
function, the specification of the Dark matter content of the universe
at all times (Figure~\ref{galpics}), provided by the Millennium
simulation: a dark matter simulation that took 3 months on a
supercomputer and that is not easily repeated. For this project we had
access to 256 processors and can parallelise the Galform calculation
into 40 sub-volumes.

%f2 #&#
\begin{figure*}

\includegraphics{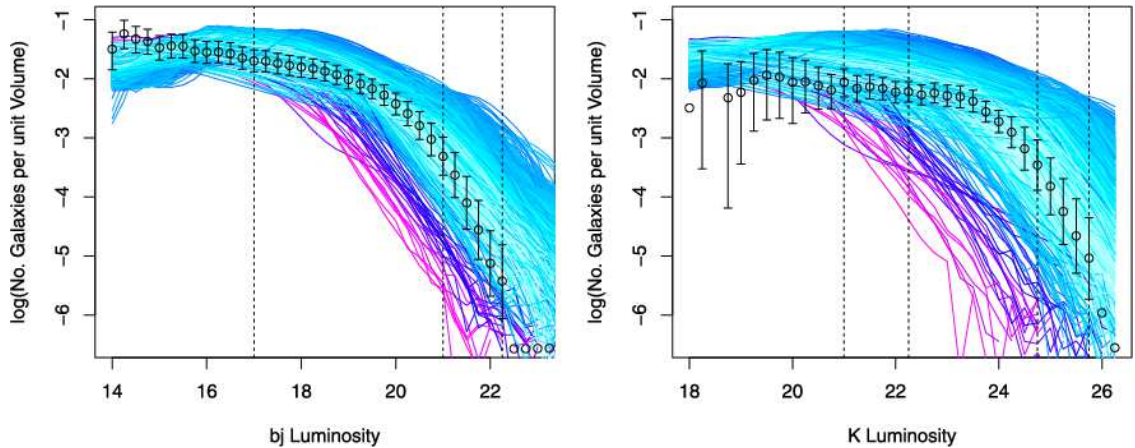}

\caption{The bj (left) and K (right) luminosity functions giving the (log)
number of galaxies per unit volume,
binned by luminosity. Black points: observed data, along with 2
sigma intervals representing all relevant uncertainties identified in
Section \protect\ref{ssecUn}. The coloured lines are the Galform
outputs from
993 wave 1 runs of the model, none
of which were found to be acceptable. The vertical lines show the 7
outputs $f(x)$
chosen for emulation (see Section \protect\ref{ssecEHM} and
Table \protect\ref{tabwaves}).}
\label{fig11outs}
\end{figure*}

Of all the outputs produced by Galform, we focus our analysis on by far
the most important: the bj and K luminosity functions, which give the
log number of blue or red (i.e., young or old) galaxies, respectively,
per unit volume, binned by luminosity
(\cite{Nord}).
These observed luminosity functions, shown as the black points in
Figure~\ref{fig11outs}, are considered to be the benchmark by which
models of galaxy formation are judged.
Models will be discarded if they do not match these luminosity
functions alone, and determining the set of input parameters that give
rise to such matches is of inherent scientific worth, as it will be
highly informative regarding the various physical processes involved in
galaxy formation. Determining if any matches even exist and, if so,
obtaining a large set of runs that match this data for use in future
analysis are major goals of the project.

%s3 #&#
\section{Bayesian History Matching}\label{secMG}

This study concerns \textit{Bayesian history matching}, to identify a
collection of input parameter choices for Galform which give acceptable
matches to certain measurements on the universe. History matching is
a common term in the oil industry, where it is used to describe the
adjustment of a model of a reservoir, by modifying the input parameter
choices, until it closely reproduces the historical production and
pressure profiles recorded in that reservoir. In Durham, we have
developed a general Bayesian approach to this problem for oil
reservoirs, expanding the use of the term from finding a single match
to searching for all such matches. A good description of this work can
be found in \citet{Craig97Pressure}. This history matching
methodology is part of the general Bayesian treatment of uncertainty
in physical systems modelled by complex computer simulators. A good
reference for this area is the website for the Managing
Uncertainty in Complex Models (MUCM) project,
\url{http://www.mucm.ac.uk}. Here, we focus on those aspects of the general
methodology that are most relevant to history matching.

We want to use the Galform simulator to reproduce the observed history
of the
physical system. Therefore, we need to consider how good the match
should be in order to be acceptable. We must recognise the limitations
of the simulator as a representation of the physical system. Our
models approximate and simplify both the properties of the system and
the physical principles used to generate the corresponding system
behaviour. Even so, the mathematical implementation is still too
complex for precise solution, and so is further simplified and
approximated. Add to this our uncertainty about initial conditions,
boundary conditions and forcing functions for the system, and it is
clear that we must assess the \textit{structural discrepancy} between
model outcomes, even if well chosen, and actual physical behaviour of
the system. Our judgements about structural discrepancy determine our
views about the quality of the match that we may achieve.

The general structure of the problem is as follows. We represent the
simulator as a vector function, taking inputs $x$ which represent
system properties, and returning outputs $f(x)$ which are intended to
correspond to certain properties, $y$, of the physical system. We have
observations $z$ on $y$. We represent the difference between $z$ and
$y$ by the relation
%
%e1 #&#
\begin{equation}
\label{obserror} z = y + e,
\end{equation}
where $e$ is the vector of random observational errors, taken to be
independent of $y$ and, typically, of each other. If $f(x)$ was a
perfect representation of the system, then we would only accept a
choice $x^*$ as representing the system if $f(x^*) = y$. Because we
can only compare $f(x^*)$ with $z$, we would therefore require the
match between $f(x^*)$ and $z$ to be probabilistically consistent with
the relation $z=f(x^*)+e$.

However, because of structural discrepancy, even if we had evaluated
an appropriate choice $f(x^*)$, we would still be uncertain about the
true system value,~$y$. If we represent this residual uncertainty by
the random structural discrepancy vector $\varepsilon$ and consider
$\varepsilon$ to be independent of $f(x^*)$, then we can write
%
%e2 #&#
\begin{equation}
\label{moddisc} y = f\bigl(x^*\bigr) + \varepsilon,
\end{equation}
where, for example, the variance of each element of $\varepsilon$
expresses our judgement about how well the corresponding element of
$f(x^*)$ is expected to reproduce that element of the system, and the
correlation between two elements of $\varepsilon$ expresses our
judgements about the similarities of the issues relating to each
component of the discrepancy. We may view $\varepsilon$ as a way of
expressing the sense that we are prepared to tolerate a less than
perfect match, and explore the effect of different choices for this
tolerance on the space of acceptable parameter matches. [For a much
more detailed treatment of the concept of model discrepancy, see
\citet{Goldstein09Reify}
and the accompanying discussion.] Specification of
beliefs for $\varepsilon$ may partly be carried out by experiments on the
simulator itself [e.g., by exploring the effect of perturbing
the forcing function or adding some internal randomness to the
propagation of an internal state vector propagated over time by the
model; see, e.g., \citet{IVassMD}]. However, a large component of
such specification comes from
the scientifically grounded but subjective judgements of the expert.

Combining (\ref{obserror}) and (\ref{moddisc}), we therefore consider the
match acceptable if it is probabilistically consistent with the
relation
%
%e3 #&#
\begin{equation}
\label{zdisc} z = f\bigl(x^*\bigr) + \varepsilon+ e.
\end{equation}
Our aim is to identify the collection, $\chi(z)$, of all choices of
$x^*$ which would give acceptable fits to historical data or, at the
least, to identify a wide range of elements of $\chi(z)$. If our input
parameter space is low dimensional, and the function is very fast to
evaluate, then we can find $\chi(z)$ by evaluating the function
everywhere and identifying the collection of all choices $x^*$
consistent with (\ref{zdisc}). However, for most complex physical
models, it is infeasible to evaluate the simulator at enough choices
to search the input space exhaustively. Therefore, we must construct a
representation of our uncertainty about the value of the simulator at
each input choice for which we have not yet evaluated the
simulator. This representation is termed an \textit{emulator}. The
emulator both suggests an approximation to the function and also
contains an assessment of the likely magnitude of the error of the
approximation. A common choice of form for emulation of component
$f_i$ is
%
%e4 #&#
\begin{equation}
\label{emulator} f_i(x) = \sum_j
\beta_{ij} g_{ij}(x_{A_i}) + u_i(x_{A_i})
+ w_i(x),
\end{equation}
where the active variables $x_{A_i}$ are subsets of the 17 inputs,
$ B = \{ \beta_{ij} \}$ are unknown scalars, $g_{ij}$ are known
deterministic functions of $x_{A_i}$, for example, polynomials,
$u_i(x_{A_i})$ is
a Gaussian process or, in a less fully specified version, a weakly
second order stationary stochastic process, with, for example,
correlation function
%
%e5 #&#
\begin{eqnarray}
\label{corr}
&&\operatorname{Corr}\bigl(u_i(x_{A_i}),u_i
\bigl(x'_{A_i}\bigr)\bigr) \nonumber\\[-8pt]\\[-8pt]
&&\quad= \exp\bigl(- \bigl\|
x_{A_i}-x_{A_i}'\bigr\|^2 /
\theta_i^2\bigr),\nonumber
\end{eqnarray}
and $w_i(x)$ is an uncorrelated nugget. $B g(x)$ expresses global
variation in $f$, while $u(x)$ expresses local
variation in $f$. We fit the emulators, given a collection of
carefully chosen simulator evaluations, using our favourite
statistical tools, guided by expert judgement. We use detailed
diagnostics to check emulator validity. A~good introduction to
function emulation is given by \citet{OHagan06Tutorial}.

Using the emulator, we can obtain, for each choice of inputs $x$, the
mean and variance, $\mathrm{E}(f(x))$ and $\operatorname{Var}(f(x))$.
Applying relation
(\ref{zdisc}), for $ x \in\chi(z)$, gives $\operatorname{Var}(z_i
-\mathrm{E}(f_i(x))) =
\operatorname{Var}(f_i(x)) + \operatorname{Var}(\varepsilon_i) +
\operatorname{Var}(e_i)$. We can therefore
calculate, for each output $f_i(x)$, the ``implausibility'' if we
consider the value $x$ to be a member of $\chi(z)$. This is the
standardised distance between $z_i$ and $\mathrm{E}(f_i(x))$, which is
%
%e6 #&#
\begin{eqnarray}
\label{Implaus} I_{(i)}^2(x)&=&\bigl|z_i - \mathrm{E}
\bigl(f_i(x)\bigr)\bigr|^2\nonumber\\[-8pt]\\[-8pt]
&&{}/\bigl[\operatorname{Var}
\bigl(f_i(x)\bigr) + \operatorname{Var}(\varepsilon_i) +
\operatorname{Var}(e_i)\bigr].\nonumber
\end{eqnarray}
Large values of $I_{(i)}(x)$ suggest that it is implausible that $x
\in\chi(z)$. The implausibility calculation can be performed
univariately, or by multivariate calculation over sub-vectors. The
implausibilities are then combined, such as by using $I_M(x)=\max_i
I_{(i)}(x)$, and can then be used to identify regions of $x$ with
large $I_M(x)$ as implausible. With this information, we can then
refocus our analysis on the ``nonimplausible'' regions of the input
space, by making more simulator runs and refitting our emulator over
such subregions and iteratively repeating the analysis. This is a
form of iterative global search aimed at finding all choices of $x$
which would give acceptable fits to historical data. We may find
$\chi(z)$ is empty, which is a strong warning of problems with our
simulator or with our data.

History matching may be compared with model calibration which aims to
identify the one ``true'' value of the input parameters $x^*$. Often,
we will prefer to carry out a history match because either we do not
believe in a unique true input value for the model or we are unsure as
to whether any good choices of input parameters exist. Further,
full probabilistic calibration analysis may be difficult, as,
typically, $\chi(z)$ will comprise a tiny volume of the original
parameter space. Therefore, even if there is an eventual intention to
carry out a full probabilistic calibration, it is often good practice
to history match first, in order to check the simulator and to reduce
the original parameter space down to $\chi(z)$.

Finally, a note on the methods used in this study. We may carry out a
full Bayes analysis, with complete joint probabilistic specification
of all of the uncertain quantities in the problem. Alternatively, we
may carry out a Bayes linear analysis, based just on a prior
specification of the means, variances and covariances of all
quantities of interest. Probability is the most common choice, but
there are advantages in working with expectations, as the uncertainty
specification is simpler and the analysis is much more technically
straightforward. Bayes linear analysis [for a detailed account, see
\citet{Goldstein07BayesLinearBookshort}] is based around these
updating equations for
mean and variance:
%
%e7 #&#
%e8 #&#
\begin{eqnarray}
\label{BLmean} \mathsf{E}_z[y] &=& \mathrm{E}(y)+\operatorname
{Cov}(y,z)\operatorname{Var}(z)^{-1}\bigl(z-\mathrm{E}(z)\bigr),
\\
\label{BLvar}\qquad
\mathsf{Var}_z[y] &= & \operatorname{Var}(y)\nonumber\\[-8pt]\\[-8pt]
&&{}-
\operatorname
{Cov}(y,z)\operatorname{Var}(z)^{-1}\operatorname{Cov}(z,y).\nonumber
\end{eqnarray}
History matching fits naturally with this approach and the Galform
study has been analysed using Bayes linear methods. There are natural
probabilistic counterparts, which we expect could have found similar
\mbox{history} matches to those we discovered, but with considerably
more effort in prior specification and computation.

%f3 #&#
\begin{figure*}

\includegraphics{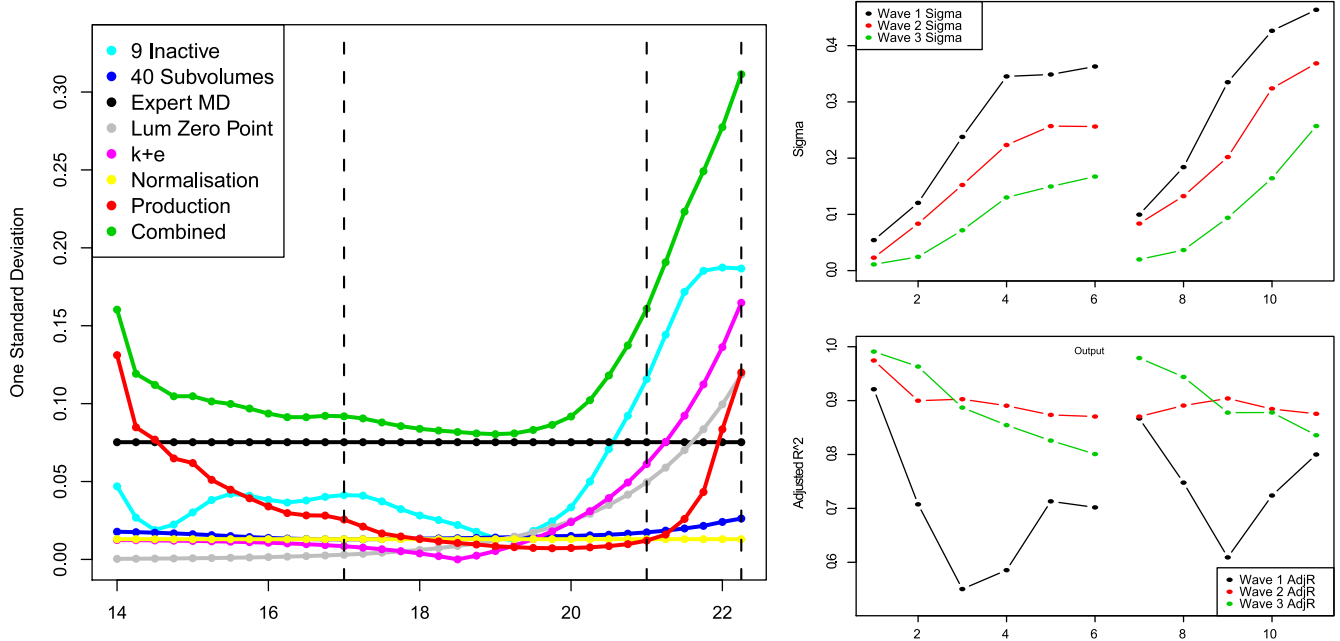}

\caption{Left panel: the sd of each contribution from the various
sources of uncertainty for the full range of the bj luminosity function
(the $x$-axis is the same as Figure \protect\ref{fig11outs}). The
vertical lines
represent the three bj outputs chosen for emulation in wave 1. Green
line: the total uncertainty due to all contributions; this value is
used for the error bars in Figure \protect\ref{fig11outs}. The K luminosity
results are similar. The residual standard deviation $\sigma$ for waves
1 to 3 (top right panel) and the adjusted R$^2$ for waves 1 to 3
(bottom right panel) for the polynomial part $\mathit{Bg}(x)$ of each emulator
[equation (\protect\ref{emulator})]. We fit high-dimensional cubic polynomials
due to having large run numbers. First 6 connected points: bj outputs
chosen for emulation, later 5 are the K outputs (shown as vertical
lines in Figure \protect\ref{figbjlum1to5}, left and right panels,
resp.). See Vernon, Goldstein and
Bower (\protect\citeyear{Vernon10CS}).} \label{figelic}
\end{figure*}

%s4 #&#
\section{Application to a Galaxy Formation Simulation}\label{secIV}

%s4.1 #&#
\subsection{Sources of Uncertainty}\label{ssecUn}

We now describe the application of the methodology introduced in
Section~\ref{secMG} to the Galform model described in Section~\ref{secRB}.
In order to determine the meaning of an acceptable match, it is
essential that we identify all sources of uncertainty that lie between
the model output $f(x)$ and reality $y$.
Note that the majority of these uncertainties have been neglected or
ignored in even the most detailed of previous analyses.
As discussed in Section~\ref{secMG}, the uncertainties separate into
two classes: the model discrepancy $\varepsilon$ and the observation
errors $e$.
In the case of Galform, the model discrepancy was decomposed into three
uncorrelated contributions $\varepsilon= \Phi_{\mathrm{IA}} + \Phi_{\mathrm{DM}} + \Phi
_E $ where:

\textit{$\Phi_{\mathrm{IA}}$ Inactive variable uncertainty}: due to coding issues,
for the first three waves we could not vary all 17 parameters
simultaneously. The 9 least active inputs were fixed and their effects
represented by this term.

\textit{$\Phi_{\mathrm{DM}}$ dark matter uncertainty}: unknown configuration of
dark matter in the universe, assessed from computer model experiments
on the 40 sub-volumes.

%t2 #&#
\begin{table*}[b]
\caption{Summary of the 4 waves of emulation. Col. 2: the no. of
model runs used to construct the emulator; col. 3: no. of outputs
emulated, col. 4: the no. of active variables; col. 5--8: the
implausibility thresholds;
col. 9: the percentage of the parameter space deemed nonimplausible}
\label{tabwaves}
\begin{tabular*}{\tablewidth}{@{\extracolsep{\fill}}lcccccccr@{}}
\hline
\textbf{Wave} & \textbf{Runs} & \textbf{Outputs emul.}
& \textbf{Active inputs} & $\bolds{I_M}$ & $\bolds{I_{2M}}$ &
$\bolds{I_{3M}}$ & $\bolds{I_{\mathit{MV}}}$ & \multicolumn{1}{c@{}}{\textbf{\% Space}}\\
\hline
1 & \hphantom{0}993 & \hphantom{1}7 & \hphantom{1}5 & --& 2.7 & 2.3 & --& 14.9\%\hphantom{00} \\
2 & 1414 & 11& \hphantom{1}8 &--& 2.7 & 2.3 & -- & 5.9\%\hphantom{00} \\
3 & 1620 & 11& \hphantom{1}8 & --& 2.7 & 2.3 & 26.75 & 1.6\%\hphantom{00} \\
4 & 2011 & 11& 10 & 3.2 & 2.7 & 2.3 & 26.75 & 0.26\%\hphantom{0} \\
5 & 2000 & -- & -- & 2.5 & -- & -- & 26.75 & 0.039\% \\
\hline
\end{tabular*}
\end{table*}

\textit{$\Phi_E$ Subjective expert assessment of model discrepancy}
between full Galform model (all 17 inputs and correct dark matter) and
real universe. Using a detailed elicitation tool, the cosmologist was
able to specify a multivariate covariance structure for $\Phi_E$
representing beliefs about the known deficiencies of the model
(over/under abundance of matter and ageing rates of red/blue galaxies),
leading to positive correlations between the discrepancy for bj outputs
and smaller positive correlations across bj and K outputs. Our methods
also incorporate a sensitivity analysis regarding the expert's
assessment of $\Phi_E$
(\cite{Goldstein09imprecise}).

The four contributions to the observation errors $e$ are as follows:

\textit{Luminosity zero point error}: correlated uncertainty across
luminosity outputs due to difficulty of defining galaxy of ``zero'' brightness.

\textit{The $k+e$ error}: a highly correlated error on all output points
due to (i) galaxies being so far away it takes light billions of years
to reach us and (ii) galaxies moving away from us so quickly their
light is redshifted.

\textit{Normalisation error}: correction for over/under population of
galaxies in local universe using theoretical considerations of universe
on large scales.

\textit{Galaxy production error}: uncertain theoretical correction due to
bright/faint galaxies being measured up to relatively large/short
distances from Earth.

Note that the observations represent \textit{theory laden data}
having been heavily preprocessed prior to our analysis\vadjust{\goodbreak} and, hence, it
would be dangerous to neglect any one of the above observational
errors. All the above uncertainties
are shown for the full bj luminosity function in Figure~\ref{figelic}
(left panel), plotted as one standard deviation against luminosity [the
$x$-axis is the same as Figure~\ref{fig11outs} (left panel)].

%f4 #&#
\begin{figure*}

\includegraphics{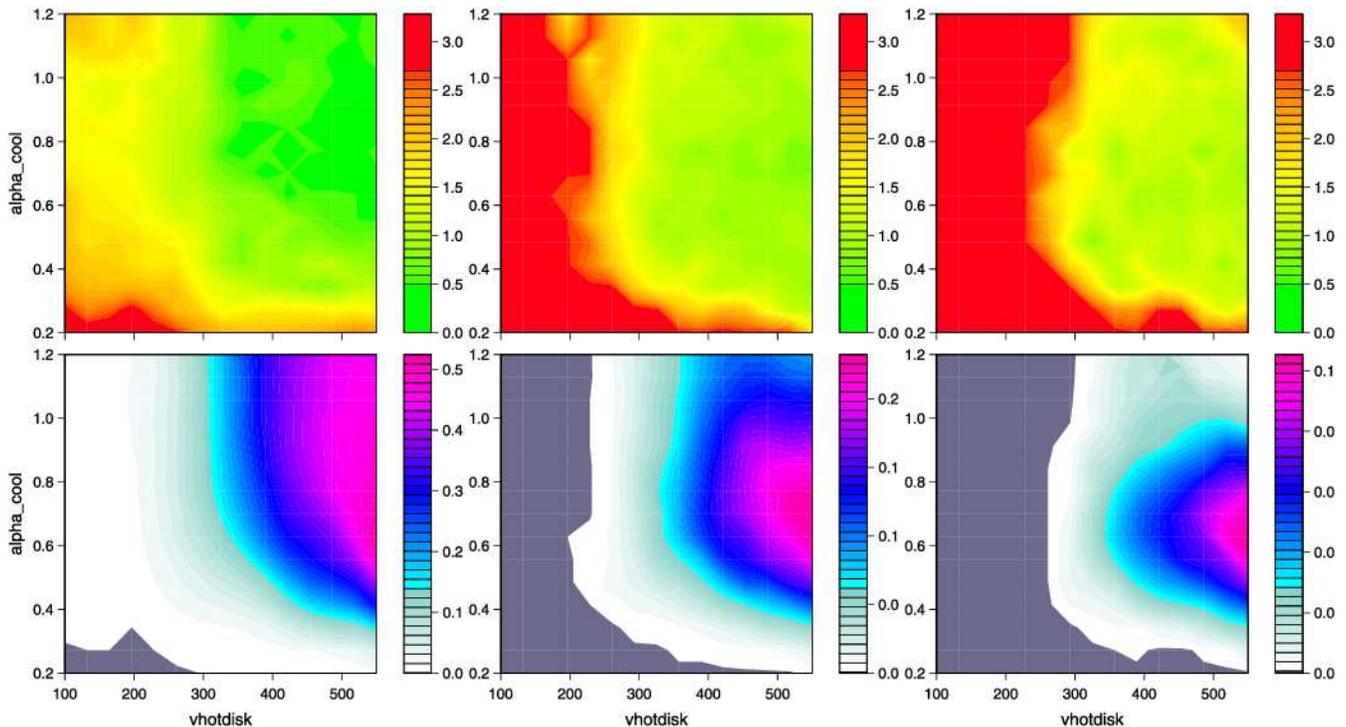}

\caption{The top three panels give waves 1, 2 and 3 minimised
implausibility projection plots in the $\vhotdisk$--$\alphacool$
plane, representing
$I_{2M}(x)$ minimised across the remaining 15 inputs. The red region
indicates high implausibility where input points will be discarded,
green/yellow: nonimplausible points.
The bottom three panels give the optical depth plots, showing the
fraction of the hidden 15-dimensional volume that satisfies the
implausibility cutoff, at that grid-point.}
\label{figW123imp}
\end{figure*}

%s4.2 #&#
\subsection{Emulation and Iterative History Matching}\label{ssecEHM}

We proceed to emulate in iterations or waves as described in
Section~\ref{secMG}. In each wave we design a space filling set of runs,
choose a subset of viable outputs $f_i(x)$ for emulation, for each
output choose a subset of active inputs $x_A$ and then construct a
Bayes linear emulator
for $f_i(x)$ using equations (\ref{emulator}) and (\ref{corr}).
The emulators are combined with the subjectively assessed model
discrepancy and the observation errors to
produce an implausibility measure $I_{(i)}(x)$ for each output
[equation (\ref{Implaus})].
We then discard regions of input space $x$ that do not satisfy cutoffs
on $I_M(x)$, $I_{2M}(x)$ and $I_{3M}(x)$ [the first, second and third
highest $I_{(i)}(x)$].
Table~\ref{tabwaves} summarises the 4 waves that were performed. For
example, in wave 1 we emulated only 7 outputs (shown as vertical dotted
lines in Figure~\ref{fig11outs}) and used only 5~active variables for
each emulator, imposing cautious implausibility constraints on only
$I_{2M}(x)$ and $I_{3M}(x)$ [as $I_M(x)$ can be
sensitive to inaccuracies in the emulators]. At each wave we performed
200 diagnostic runs to check emulator performance.

%f5 #&#
\begin{figure*}

\includegraphics{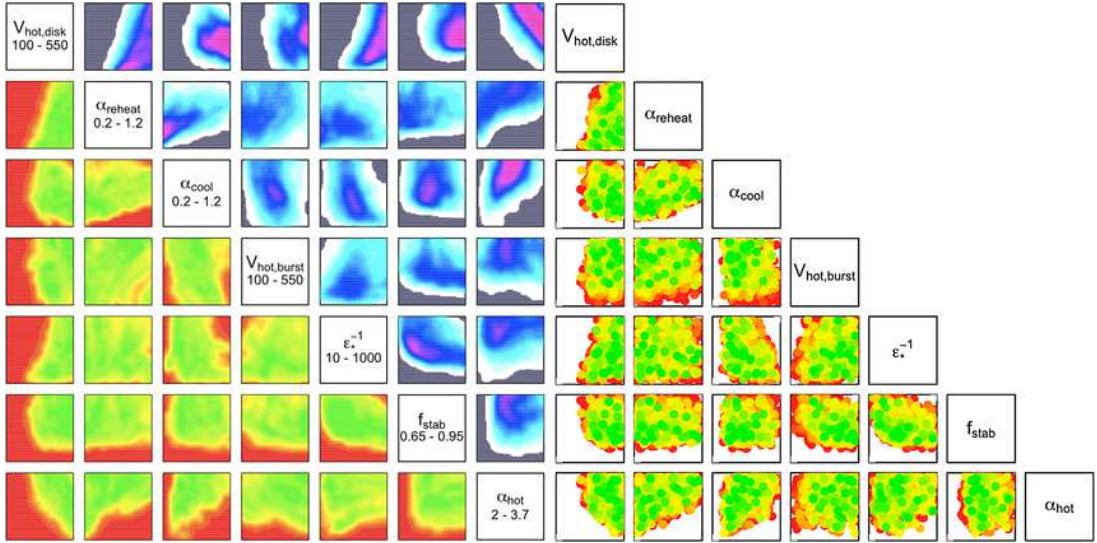}

\caption{Left: wave 4 minimised implausibility (below diagonal) and
optical depth (above diagonal) projections (see
Figure~\protect\ref{figW123imp}). Right: the wave 5 runs coloured by the
implausibility consistent with the left panel, now with no emulator
uncertainty, confirming the wave 4 predictions.}\label{figPairsW4}
\end{figure*}

In each new wave we perform more runs, the emulators become more
accurate, the implausibility measures more informative and, hence, we
are able to discard more space as implausible than in the previous
wave. Explicit improvement in the emulators over the first three waves
is shown in Figure~\ref{figelic} (top right and bottom right panels).
We expect this emulator improvement, as at each wave (a) there are a
higher density of runs which improves the Gaussian process part of the
emulator, (b) we can choose more active inputs $x_A$, (c) we are
emulating a smoother function since it is defined over a smaller volume
and (d) we can hence choose more outputs to emulate.
The iterative nature of the space reduction process is\vadjust{\goodbreak} the main reason
the history matching approach is so powerful and is shown in
Figure~\ref{figW123imp} for waves 1 to 3. The percentage of input
space remaining is given in Table~\ref{tabwaves}.

%s4.3 #&#
\subsection{Iterative History Matching: Waves 4 and 5 Results}

We performed 4 waves of history matching in order to identify the set
of input parameters consistent with the luminosity function observations.
Various 2D projections of the nonimplausible set of inputs at wave 4
are shown in Figure~\ref{figPairsW4} (left panel), where the
projections are onto the
subspaces of pairs of 7 of the most interesting input parameters, out
of the full 17 given in Table~\ref{tabinputs}. These projections,
along with higher-dimensional equivalents, provide the cosmologists
with detailed insight into to the behaviour of the Galform model:
indeed, there was much initial surprise as to the extent of the nonimplausible
region in some directions, despite it
occupying a tiny percentage of the original input space of only $0.039
\%$.

%f6 #&#
\begin{figure*}[b]
\vspace*{6pt}
\includegraphics{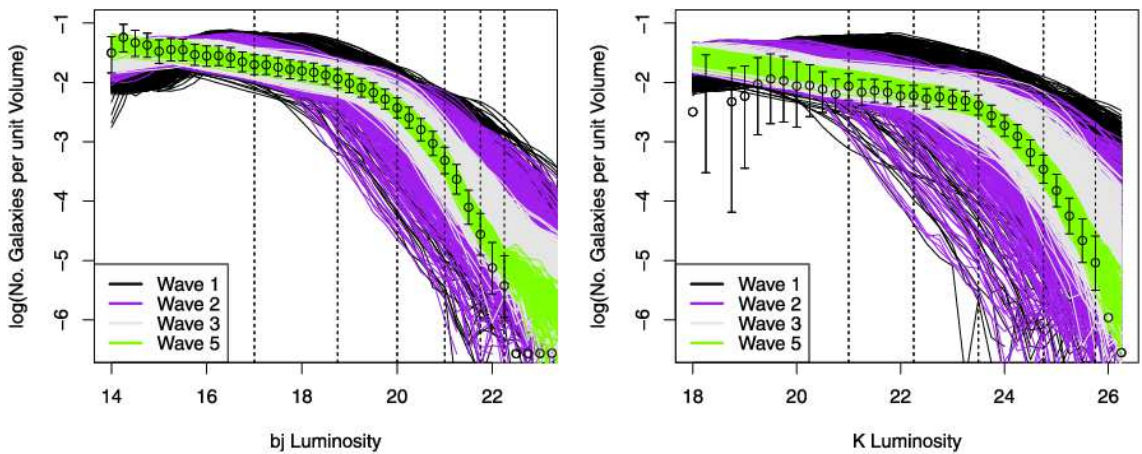}

\caption{Left: the bj luminosity function output for the first 500
runs of waves 1, 2 and 3 and the wave 5 acceptable runs that satisfy
$I_M(x)<2.5$. Right: K luminosity. The disparity at luminosity $\le$
19 between K luminosity data and wave 5 runs is due to
the limited resolution of the dark matter simulation [see Bower et~al.
(\protect\citeyear{B06})] and so is not considered of interest.}
\label{figbjlum1to5}
\end{figure*}

As the wave 4 emulator variances were smaller than the combined model
discrepancy and observation errors, the iterations were terminated. A final
set of wave 5 runs was generated both to confirm the predictions made
by the wave 4 emulator
of the extent of the region of acceptable matches and to obtain a large
set of acceptable runs for use by the cosmologists, a major goal of the
project. These wave 5 runs are shown in Figure~\ref{figPairsW4} (right
panel) with the same implausibility colour scale as in the left panel,
but now without any emulator uncertainty. Large numbers of acceptable
runs were found, and 306 runs were found to satisfy the more strict
cutoff $I_M(x)< 2.5$, superior to any matches previously found by the
cosmologists. The
outputs of these acceptable runs, along with those of previous waves,
are shown in Figure~\ref{figbjlum1to5}. Note that the acceptable runs
are good matches across all luminosities, not just at the 11 ouptuts
chosen for emulation.

%s5 #&#
\section{Conclusion}

The task of finding matches between complex galaxy formation
simulation output and observations of the real universe
represents a fundamental challenge within cosmology. Even to
define what we mean by an acceptable match requires an assessment
of model discrepancy, which can only come through a, necessarily
subjective, scientific judgement based on many years of
experience in constructing such simulations. Therefore, this
problem fits naturally into a Bayesian framework, in which we
treat all of the uncertainties arising from properties of the
simulator or of the data in a unified manner.

The resulting problem, of identifying matches consistent with our
uncertainty measures, is extremely challenging, involving
understanding the simulator's behaviour over a high-dimensional
input parameter space. It is difficult to see how to proceed
without the use of carefully constructed Bayesian emulators that
represent our beliefs about the behaviour of the deterministic
function at all points in the input space and which are fast to
evaluate. These emulators are used within an iterative history
matching strategy that seeks only to emulate in detail the most
interesting parts of the input space, and thus provides a global
search algorithm which gives a practical and tractable Bayesian
solution to the problem.

We have demonstrated this solution for the galaxy formation
simulator. Specifically, we have identified the regions of input
space of interest, occupying $0.039\%$ of the initial volume, and
provided the cosmologists with a large set
of runs that yield acceptable matches: a major goal of the
project. An account of the impact of this approach within cosmology is
given in \citet{Bower09ParaGalf2}. A history match is in most
cases sufficient for the scientists' needs, both for model
analysis and development. However, even if a more detailed, fully
probabilistic Bayesian analysis is required, perhaps of a well-tested
and highly accurate model, a~history match is usually a
good precursor to the calibration exercise, to rule out the vast
areas of input space that would possess extremely low posterior
probability.

\section*{Acknowledgements}

Supported by EPSRC, STFC, MUCM Basic Technology initiative.

%suskaldyti doi

% imsref loaded by lrinkeviciute, 2013-06-18 13:25:18
% imsref loaded by lrinkeviciute, 2013-06-18 13:31:36
% imsref loaded by lrinkeviciute, 2013-06-18 13:34:12
%

\end{document}